# New shapes of light-cone distributions of the transversely polarized $\rho$-mesons

A. P. Bakulev[†], S. V. Mikhailov[‡]

*Bogolyubov Lab. of Theoretical Physics,*
*Joint Institute for Nuclear Research,*
*141980, Moscow Region, Dubna, Russia*

**Abstract**

The leading twist light-cone distributions for transversely polarized $\rho$-, $\rho'$- and $b_1$-mesons are re-analyzed in the framework of QCD sum rules with nonlocal condensates. Using different kinds of sum rules to obtain reliable predictions, we estimate the 2-, 4-, 6- and 8-th moments for transversely polarized $\rho$- and $\rho'$-meson distributions and re-estimate tensor couplings $f^T_{\rho,\rho',b_1}$. We stress that the results of standard sum rules also support our estimation of the second moment of the transversely polarized $\rho$-meson distribution. New models for light-cone distributions of these mesons are constructed. Phenomenological consequences from these distributions are briefly discussed. Our results are compared with those found by Ball and Braun (1996), and the latter is shown to be incomplete.

## 1 Introduction

In this paper, we complete our investigation of the leading twist light-cone distribution amplitudes (DAs) for lightest transversely polarized mesons with quantum numbers $J^{PC} = 1^{--}$ ($\rho_\perp$, $\rho'_\perp$), $1^{+-}$ ($b_{1\perp}$) in the framework of QCD sum rules (SRs) with nonlocal condensates (NLC). These DAs are important ingredients of the "factorization" formalism [1] for any hard exclusive reactions involving $\rho$-mesons. For this reason, the DAs have been attractive for theorists for a long time: the main points are presented in [2, 3], a detailed revised version of the standard approach is in [4], and a generalization to the next twists is in [5]. The leading twist DA $\varphi^T_{\rho,\rho',b_1}(x,\mu^2)$ parameterizes the matrix elements of the tensor current with transversely polarized $\rho(770)$- and $\rho'(1465)$-mesons ($J^{PC} = 1^{--}$)

$$\langle 0 \mid \bar{u}(z)\sigma_{\mu\nu}d(0) \mid \rho_\perp(p,\lambda)\rangle\Big|_{z^2=0} = if^T_{\rho_\perp} \left(\varepsilon_\mu(p,\lambda)p_\nu - \varepsilon_\nu(p,\lambda)p_\mu\right) \int_0^1 dx \; e^{ix(zp)} \; \varphi^T_{\rho_\perp}(x,\mu^2) + \ldots, \quad (1)$$

and the $b_1(1235)$-meson ($J^{PC} = 1^{+-}$)

$$\langle 0 \mid \bar{u}(z)\sigma_{\mu\nu}d(0) \mid b_1^+(p,\lambda)\rangle\Big|_{z^2=0} = f^T_{b_1}\epsilon_{\mu\nu\alpha\beta}\varepsilon^\alpha(p,\lambda)p^\beta \int_0^1 dx \; e^{ix(zp)} \; \varphi^T_{b_1}(x,\mu^2) + \ldots \quad (2)$$

(here dots represent higher-twist contributions, explicitly defined in Appendix A, see Eqs.(A.9)-(A.10) and ref.[5]). In the above definitions, $p_\nu$ and $\varepsilon_\mu(p,\lambda)$ are the momentum and the polarization vector of a meson, respectively, and $\mu^2$ is normalization point.

[†]E-mail: bakulev@thsun1.jinr.ru
[‡]E-mail: mikhs@thsun1.jinr.ru

In the framework of the standard approach, one should restrict oneself to an estimate of the second moment $\langle \xi^2 \rangle$ of the DA to restore its shape[1]. In other words, the variety of different DA shapes is reduced to the 1-parameter family of "admissible" DAs: $\varphi(x; a_2) = 6x(1-x)\left[1 + a_2 C_2^{3/2}(2x-1)\right]$. This family includes both the asymptotic DA ($a_2 = 0$) and Chernyak–Zhitnitsky model [2] for the pion DA ($a_2^{\pi|\text{CZ}} = -2/3$). For the pion case, one can think it is rather enough: most of debates (see [2, 6, 10, 11, 12] and refs. therein) about the shape of this DA are concerned just with the value of coefficient $a_2$ – is it close to 0 or to $a_2^{\pi|\text{CZ}}$? In our opinion, advocated since 1986 [6], the shape of the pion DA is not far from the asymptotic one [6, 7, 13, 14, 8]. Only recently, researchers have tried to extract the next Gegenbauer coefficient [12] and other parameters of the pion DA [15] from experimental data. But, in general, there is no principle to exclude a more rich structure for a hadron DA. In this case, the standard approach is definitely out of its applicability range, and one should use more refined techniques, e.g., the QCD SRs with NLC.

This work was started in [8] where the "mixed parity" NLC SR for DAs of $\rho$- and $b_1$-mesons, the particles possessing different P-parity, was analyzed. We concluded that, to obtain a reliable result, one should reduce model uncertainties due to the nonlocal gluon contribution. Separate SRs for each P-parity channel should be preferable for this purpose, and here we construct these "pure parity" SRs for corresponding DAs. The SR of this type possesses a low sensitivity to the gluon model but involves contributions from higher twists[2]. To construct a refined "pure parity" SR for twist 2 DA, one must resolve the corresponding system of equations (see Appendix A). We realize this solution using the duality transformation, introduced in our previous work [16]. The negative parity NLC SR for the transversely polarized $\rho$-, $\rho'$-mesons works rather well and allows us to estimate the 2-nd, 4-th, 6-th, and 8-th moments of the leading twist DAs. The positive parity SR for the transversely polarized $b_1$-meson can provide only the value of the $b_1$-meson tensor coupling, $f_{b_1}^T$. We suggest the models for these DAs and check their self-consistency, based upon both "pure" and "mixed" NLC SR. The DA shape $\varphi_{\rho_\perp}^T(x)$ differs noticeably from the known one. Finally, we inspect how these models can influence the $B \to \rho e \nu$ decay form factors.

The approach has been grounded in [6, 7, 17], the calculation technique is the same as in [7, 8]; therefore, the corresponding details are omitted below. Some important features of the NLC SRs approach would be briefly recalled. The original tools of NLC SR are nonlocal objects like $M_S(z^2) = \langle \bar{q}(0) E(0, z) q(z) \rangle$ [3] or $M_V^\mu(z) = \langle \bar{q}(0) \gamma^\mu E(0, z) q(z) \rangle$, rather than constant quantities of $\langle \bar{q}(0) q(0) \rangle$-type. Note that, in deriving sum rules, one can always make a Wick rotation and treat all the coordinates as Euclidean, $z^2 = -\tilde{z}_{Eucl}^2 < 0$. NLC $M_S(z^2)$ can be expanded in the Taylor series over the standard (local) condensates, $\langle \bar{q}(0) q(0) \rangle$, $\langle \bar{q}(0) \nabla^2 q(0) \rangle$, and over "higher dimensions" (see details of the expansion of different NLCs in [18]),

$$M_S(z^2) = \langle \bar{q}(0) q(0) \rangle - \frac{\tilde{z}^2}{8} \langle \bar{q}(0) \nabla^2 q(0) \rangle + \ldots. \qquad (3)$$

So, one can return to the standard SR by truncating this series. But, in virtue of the cut off, one loses an important physical property of nonperturbative vacuum – the possibility of vacuum quarks (gluons) to flow through vacuum with a nonzero momentum $k_{q(g)} \neq 0$. The parameter $\langle k_q^2 \rangle$, fixing the average virtuality of vacuum quarks, can be interpreted as a measure of condensate "nonlocality" $\lambda_q^2$,

$$\langle k_q^2 \rangle = \lambda_q^2 = \frac{\langle \bar{q}(0) \nabla^2 q(0) \rangle}{\langle \bar{q}(0) q(0) \rangle} = \frac{\langle \bar{q}(0) \left(ig\sigma_{\mu\nu} G^{\mu\nu}\right) q(0) \rangle}{2 \langle \bar{q}(0) q(0) \rangle} \quad \text{[chiral limit]}$$

---

[1] We should note in this respect that the standard approach could not provide a reliable estimate even for the second moment of DA, see [6, 7, 8, 9]

[2] as was noted in [4].

[3] Here $E(0, z) = P \exp(i \int_0^z dt_\mu A_\mu^a(t) \tau_a)$ is the Schwinger phase factor required for gauge invariance.

The $\lambda_q^2$ was estimated from the mixed condensate of dimension 5, $\lambda_q^2 \approx 0.4 - 0.5$ GeV$^2$ [19, 20]. It is important that its value is of an order of the characteristic hadronic scale, $\lambda_q^2 \sim m_\rho^2 \approx 0.6$ GeV$^2$, therefore the nonlocality effect can be large, and it should be taken into account in QCD SR. Really, the second term in the expansion (3) of $M_S(z^2)$ that is inverse of the first one in sign becomes of an order of the first term at $|z^2| \sim 1/m_\rho^2$ due to the estimate $|\lambda_q^2 z^2| \sim 1$. Moreover, we should take into account the whole set of $\left(\lambda_q^2 z^2\right)^n$-type corrections, appearing in the Taylor expansion. These corrections just form the decay rate of the NLC ($M_S(z^2)$) in the main. The sensitivity to this rate is crucial for the DA moment SR: it leads to much softer behavior of DA near the end points $x = 0, 1$ and allows one to extend QCD SR to higher moments $\langle \xi^N \rangle \equiv \int_0^1 \varphi(x)(2x-1)^N dx$, as it was shown in [6].

Since neither QCD vacuum theory exists yet, nor higher dimension condensates are estimated, it is clear that merely the models of NLC can be suggested (Appendix B). Here we apply the simplest ansatz to NLC [7, 8] that takes into account only the main effect $\langle k_q^2 \rangle = \lambda_q^2 \neq 0$ and fixes a length of the quark-gluon correlations in QCD vacuum $\Lambda = 1/\lambda_q \approx 0.8$(Fm) [6, 7]. This suggestion leads to the simple Gaussian decay for $M_S(z^2)$, while the coordinate behavior of other NLCs looks more complicated. Certainly, the single scale of decay for all types of NLC (see Appendix B) looks as a crude model. But, the model can be rather crude if one deals with SRs only for the first few moments $\langle \xi^N \rangle$, because for these integral characteristics the details of NLC behavior appears to be not very important (see discussion in sect.5). An alternative case is provided by a special SR [13, 14] constructed directly for the shape of DAs.

In nowadays, the lattice calculations of NLC provide an inspiring knowledge [21, 22] for QCD SR. The latter measurement in [22] confirms the validity of the Gaussian ansatz for $M_S(z^2)$ (at a small distance) as well as the value of the parameter $\lambda_q^2$.

## 2 "Duality" transformation

To obtain sum rule, we start with a 2-point correlator $\Pi^{\mu\nu;\alpha\beta}(q)$ of tensor currents $J_{(N)}^{\mu\nu}(x) = \bar{u}(x)\sigma^{\mu\nu}(z\nabla)^N d(x)$ ($z$ is a light-like vector, $z^2 = 0$),

$$\Pi_{(N)}^{\mu\nu;\alpha\beta}(q) = i \int d^4x \ e^{iq\cdot x} \langle 0|T\left[J_{(0)}^{\mu\nu+}(x) J_{(N)}^{\alpha\beta}(0)\right]|0\rangle \quad (4)$$

whose properties were partially analyzed in [3, 4, 16]. It is well known that the correlator at $N = 0$ can be decomposed in invariant form factors $\Pi_\pm$, [3, 4]

$$\Pi_{(0)}^{\mu\nu;\alpha\beta}(q) = \Pi_-(q^2)P_1^{\mu\nu;\alpha\beta} + \Pi_+(q^2)P_2^{\mu\nu;\alpha\beta} \quad (5)$$

where the projectors $P_{1,2}$, obeying the projector-type relations

$$(P_i \cdot P_j)^{\mu\nu;\alpha\beta} \equiv P_i^{\mu\nu;\sigma\tau} P_j^{\sigma\tau;\alpha\beta} = \delta_{ij} P_i^{\mu\nu;\alpha\beta} \text{ (no sum over } i\text{)}, \ P_i^{\mu\nu;\mu\nu} = 3, \quad (6)$$

are presented in Appendix A. For the general case $N \neq 0$, a similar decomposition involves 4 new independent tensors $Q_i$; they appear due to a new vector $z^\alpha$ introduced into the composite tensor current operator,

$$\begin{aligned}\Pi_{(N)}^{\mu\nu;\alpha\beta}(q) &= \Pi_-(q^2, qz)P_1^{\mu\nu;\alpha\beta} + \Pi_+(q^2, qz)P_2^{\mu\nu;\alpha\beta} + K_1(q^2, qz)Q_1^{\mu\nu;\alpha\beta} \\ &+ K_3(q^2, qz)Q_3^{\mu\nu;\alpha\beta} + K_z(q^2, qz)Q_z^{\mu\nu;\alpha\beta} + K_q(q^2, qz)Q_q^{\mu\nu;\alpha\beta} \ . \end{aligned} \quad (7)$$

Contributions of DAs, defined in Eqs.(A.9)-(A.10), to different tensor structures in decomposition (7) are mixed, see Eqs.(A.11)-(A.12). The most effective way to disentangle them in practical OPE calculations is to use explicit properties of different OPE terms under the duality transformation $\hat{D}$ (introduced in our previous work [16]) mapping any rank-4 tensor $T^{\mu\nu;\alpha\beta}$ to another rank-4 tensor $T_D^{\mu\nu;\alpha\beta} = (\hat{D}T)^{\mu\nu;\alpha\beta}$ with

$$D^{\mu\nu;\alpha\beta}_{\mu'\nu';\alpha'\beta'} = \frac{-1}{4}\epsilon^{\mu\nu}{}_{\mu'\nu'}\epsilon_{\alpha'\beta'}{}^{\alpha\beta} \text{ and } \hat{D}^2 = 1. \tag{8}$$

Our projectors $P_1^{\mu\nu;\alpha\beta}$, $P_2^{\mu\nu;\alpha\beta}$ $Q_1^{\mu\nu;\alpha\beta}$, $Q_3^{\mu\nu;\alpha\beta}$, $Q_z^{\mu\nu;\alpha\beta}$, and $Q_q^{\mu\nu;\alpha\beta}$ transform into each other under the action of $\hat{D}$:

$$\left(\hat{D}P_1\right)^{\mu\nu;\alpha\beta} = P_2^{\mu\nu;\alpha\beta} \;;\quad \left(\hat{D}Q_1\right)^{\mu\nu;\alpha\beta} = [P_1 + P_2 - Q_3]^{\mu\nu;\alpha\beta} \;; \tag{9}$$

$$\left(\hat{D}P_2\right)^{\mu\nu;\alpha\beta} = P_1^{\mu\nu;\alpha\beta} \;;\quad \left(\hat{D}Q_3\right)^{\mu\nu;\alpha\beta} = [P_1 + P_2 - Q_1]^{\mu\nu;\alpha\beta} \;; \tag{10}$$

$$\left(\hat{D}Q_z\right)^{\mu\nu;\alpha\beta} = -Q_z^{\mu\nu;\alpha\beta};\quad \left(\hat{D}Q_q\right)^{\mu\nu;\alpha\beta} = [Q_q - Q_z + Q_1 + Q_3 - P_1 - P_2]^{\mu\nu;\alpha\beta} \;. \tag{11}$$

We have shown in [16] that all terms in OPE could be divided into two classes, self-dual ($\hat{D}X_{\rm SD} = X_{\rm SD}$) and anti-self-dual ($\hat{D}X_{\rm ASD} = -X_{\rm ASD}$). For example, the perturbative term is of ASD type, whereas the 4-quark scalar condensate contribution to OPE is of SD type.

Below we introduce the shorthand notation for contributions of DAs to decomposition (7): $v_0$, $v_1$, and $v_2$ stand for $1^{--}$ ($\rho_\perp$, $\rho'_\perp$); and $u_0$, $u_1$, and $u_2$, for $1^{+-}$ ($b_1$), see Appendix A for details. For SD parts of OPE $u_i = -v_i$, and the system of equations simplifies to:

$$\frac{\Pi_\mp(q^2,qz)}{2(qz)^N q^2} = \mp v_0 - v_1 - v_2 \;;\quad \frac{K_{1,3}(q^2,qz)}{2(qz)^N q^2} = \mp v_1 + v_2 \;;\quad \frac{K_q(q^2,qz)(=-2K_z(q^2,qz))}{4(qz)^N q^2} = v_2 \;, \tag{12}$$

whereas for ASD parts $u_i = v_i$, and we have:

$$\frac{\Pi_\mp(q^2,qz)}{2(qz)^N q^2} = \mp v_0 + v_1 + v_2 \;;\quad \frac{K_{1,3}(q^2,qz)}{2(qz)^N q^2} = -v_1 - v_2 \;;\quad \frac{K_z(q^2,qz)}{2(qz)^N q^2} = +v_2 \;;\; K_q(q^2,qz) = 0 \;. \tag{13}$$

By these formulas, it is possible to determine $\rho$- and $b_1$-meson DA contributions of leading and higher twists.

## 3 The "mixed parity" sum rule

The usual way [2, 4] to extract the moments of the function $\varphi^T(x)$ appeals to a correlator $J_{(N,0)}(q^2)$ of currents $J_{(N)}^{\mu\alpha}(0)z^\alpha$ and $J_{(0)}^{\mu\beta}(x)z^\beta$ defined as

$$-2i^n (zq)^{N+2} J_{(N,0)}(q^2) \equiv \Pi_{(N)}^{\mu\nu;\alpha\beta}(q)\left(z^\nu z^\beta g^{\mu\alpha}\right) = \frac{\Pi_-(q^2) - \Pi_+(q^2)}{q^2}(qz)^2, \tag{14}$$

the latter equality in (14) follows from (7) and Eqs.(A.7) in Appendix A. This correlator contains the contributions from states with different parity, $\Pi_-(q^2)$ and $\Pi_+(q^2)$ (see the analysis in [4]), therefore, the contamination from $b_1$-meson $\left(J^{PC} = 1^{+-}\right)$ in the phenomenological part of the corresponding SR is mandatory. The contamination makes it difficult to reliably extract the meson characteristics from this "mixed" SR.

The main feature of the theoretical part of $J_{(N,0)}(q^2)$ is the cancellation of the self-dual part, represented by the four-quark condensate, in the anti-self-dual expression (14). The remaining "condensate" parts of Eq. (14) contain, after the Borel transformation, the same 5 universal elements $\Delta\Phi_\Gamma(x;M^2)$ as for the $\rho^L$-, $\pi$-cases and, besides, an additional gluon contribution $\Delta\Phi'_G(x;M^2)$ (see

Appendix B). This term affects the values of moments rather strong, as was shown in [8]. The contributions from the different kinds of NLC, $\Delta\Phi_\Gamma(x;M^2)$, are symbolically noted in the r.h.s. of SR (15). So, here we get rid of the four-quark condensate that is not known very well due to a possible vacuum dominance violation. But, the price we pay for it is a high sensitivity to an ill-known gluon contribution $\Delta\Phi'_G(x;M^2)$.

The method of calculation of the NLC contributions $\Delta\Phi_\Gamma(x;M^2)$ to the theoretical part of SR is described in [6, 7, 8]. The corrected final results of the calculation are presented in Appendix B that contains all the needed explicit expressions of $\Delta\Phi_\Gamma(x;M^2)$ for the simplest physically motivated Gaussian ansatz. The final SR including DAs of $\rho$-meson and next resonances $\rho'$ and $b_1$ into the phenomenological (left) part is as follows:

$$\left(f_\rho^T\right)^2 \varphi_\rho^T(x) e^{-m_\rho^2/M^2} + (\rho \to \rho') + \left(f_{b_1}^T\right)^2 \varphi_{b_1}^T(x) e^{-m_{b_1}^2/M^2} = \int_0^{s_b^T} \rho_T^{\text{mixed}}\left(x,s;s_\rho^T,s_b^T\right) e^{-s/M^2} ds$$
$$+ \Delta\Phi_G(x;M^2) + \Delta\Phi'_G(x;M^2) + \Delta\Phi_V(x;M^2) + \Delta\Phi_T(x;M^2) \, , \quad (15)$$

where $s_\rho^T$ and $s_b^T$ are the effective continuum thresholds in $\rho$- and $b_1$-channels. Recall again that the variation of the ill-known part of gluon contribution $\Delta\Phi'_G(x;M^2)$ can reduce the second moment significantly [8]. In that paper, we suggest the following naive model: instead of the constant contribution $\Delta\varphi'_G\left(x;M^2\right) \equiv \langle\alpha_s GG\rangle/(6\pi M^2)$ (as in the standard approach), we put

$$\Delta\Phi'_G\left(x;M^2\right) = \Delta\varphi'_G\left(x;M^2\right) \frac{\theta\left(\Delta < x\right)\theta\left(x < 1-\Delta\right)}{1-2\Delta}.$$

This simulation eliminates end-point $(x = 0, 1)$ effects due to the influence of the vacuum gluon nonlocality inspired by the analysis in [17] and our experience in the nonlocal quark case. The corresponding SR leads to estimate $\langle\xi^2\rangle_\rho^T = 0.329(11)$ (see Fig.2(a) ). However, this value drastically changes, $\langle\xi^2\rangle_\rho^T \to 0.231(8)$, if we take the local expression $\Delta\varphi'_G\left(x,M^2\right)$ unchanged. Therefore, the estimate $\langle\xi^2\rangle_\rho^T = 0.329$ contains a significant model uncertainty, and the real value seems to be smaller.

Which prediction for this quantity can be obtained within the standard QCD SR approach? As one can see from Fig.2(b) (long-dashed line), the value of $\langle\xi^2\rangle_\rho^T$ cannot be estimated with a reasonable accuracy, because the standard SR **does not have real stability**. Nevertheless, the authors of [4] bravely deduce an estimate $\langle\xi^2\rangle_{\rho\ [B\&B]}^T = 0.27(4)$. We discuss this attempt in comparison with processing other SRs in greater detail in section 5.

## 4 The "pure parity" sum rules

Using the approach of Section 2, we calculate OPE terms for $\Pi_\mp$, $K_{1,3}$, and $K_{z,q}$ correlators and extract the contributions to DAs of the $\rho$- and $b_1$-mesons. This allows us to write down the SRs for DAs of the $\rho$- and $b_1$-mesons separately:

$$\left(m_\rho f_\rho^T\right)^2 \varphi_\rho^T(x) e^{-m_\rho^2/M^2} + \left(m_{\rho'} f_{\rho'}^T\right)^2 \varphi_{\rho'}^T(x) e^{-m_{\rho'}^2/M^2} = \frac{1}{2}\int_0^{s_\rho^T} \rho_T^{\text{pert}}(x;s)\, s\, e^{-s/M^2} ds$$
$$+ \Delta\tilde{\Phi}_G(x;M^2) + \Delta\tilde{\Phi}_S(x;M^2) + \Delta\tilde{\Phi}_V(x;M^2) + \Delta\tilde{\Phi}_T(x;M^2) \, ; \quad (16)$$

$$\left(m_{b_1} f_{b_1}^T\right)^2 \varphi_{b_1}^T(x) e^{-m_{b_1}^2/M^2} = \frac{1}{2}\int_0^{s_b^T} \rho_T^{\text{pert}}(x;s)\, s\, e^{-s/M^2} ds$$
$$+ \Delta\tilde{\Phi}_G(x;M^2) - \Delta\tilde{\Phi}_S(x;M^2) + \Delta\tilde{\Phi}_V(x;M^2) + \Delta\tilde{\Phi}_T(x;M^2) \, . \quad (17)$$

where $s^T_{\rho;b}$ are the effective continuum thresholds in the $\rho$- and the $b_1$-meson cases, respectively. The perturbative spectral density $\rho^{pert}_T(x;s)$ is presented in an order of $O(\alpha_s)$ in [4, 8] (Appendix B). Here we also define "tilded" functions

$$\Delta\tilde{\Phi}_\Gamma(x;M^2) \equiv \frac{1}{2}M^4\partial_{M^2}\Delta\Phi_\Gamma(x;M^2) , \qquad (18)$$

and the whole tensor NLC contribution

$$\Delta\tilde{\Phi}_T(x;M^2) \equiv \Delta\tilde{\Phi}_{T_1}(x;M^2) + \Delta\tilde{\Phi}_{T_2}(x;M^2) - \Delta\tilde{\Phi}_{T_3}(x;M^2). \qquad (19)$$

The later noticeably differs from the case of longitudinally polarized $\rho$-meson due to the opposite sign of $T_3$-term, cf. [8]. The theoretical "condensate" part in (16)-(17) contains 5 elements obtained from (18) with the same $\Delta\Phi_\Gamma(x;M^2)$ as for the $\rho^L$-meson case, whereas the self-dual four-quark contribution $\Delta\tilde{\Phi}_S(x;M^2)$ is a new element of the analysis. Note, just this self-dual part $\Delta\tilde{\Phi}_S(x;M^2)$, entering in the SRs (16) and (17) with different sign, provides the different properties of the $\rho$- and $b_1$-mesons [16].

For better understanding of the SR properties it is instructive to reduce them to standard version for $\langle\xi^N\rangle$-moments. To this end, let us take the limits $\lambda^2_q \to 0$, $\Delta\Phi_\Gamma(x,M^2) \to \Delta\varphi_\Gamma(x,M^2)$ in eqs.(16)-(17) and integrate in $x$ with weights $(1-2x)^N$ to obtain the local limit version of moment SR:

$$\left(m_\rho f^T_\rho\right)^2 \langle\xi^N\rangle^T_\rho e^{-m^2_\rho/M^2} = \qquad (20)$$
$$\frac{1}{2}\int_0^{s^T_\rho} \rho^{\text{pert}}_T(x;s)\, s\, e^{-s/M^2} ds - \frac{\langle\alpha_s GG\rangle}{24\pi}\left(\frac{N-1}{N+1}\right) - \frac{16\pi}{81}\frac{\langle\sqrt{\alpha_s}\bar{q}(0)q(0)\rangle^2}{M^2}(4N-13) .$$

This SR demonstrates a considerably lower sensitivity to the gluon condensate contribution: the gluon part does not depend on the Borel parameter $M^2$ at all, and its relative value is 6 times as low as that in the "mixed" SR. The r.h.s. of Eq. (20) is reduced at $N=0$ to the known expression, see [4], that is not sensitive to the $\rho'$ contribution, while its nonlocal version analyzed in [16] makes it possible to analyse the $\rho'$ meson. For $N > 0$, the SR is unstable due to the effect of radiative corrections, and to obtain the moment estimates, we should return to the nonlocal version, Eq.(16).

But the price one pays for this is high, the fidelity windows of the SRs are significantly reduced. For the $\rho$-meson case, fidelity windows of the Borel parameters $M^2$ shrink to $M^2 = 0.7 - 1.15$ GeV$^2$ (to be compared with $M^2 = 0.75 - 2.25$ GeV$^2$ in "mixed" SR) and demand one to take into account the $\rho'$-meson explicitly. Here we cannot obtain the $\rho'$-meson mass from SR (16) because of the enhanced perturbative spectral density ($\sim s$; this means that the differentiated SR has a spectral density $\sim s^2$ and presumably, is not stable at all); instead, we use the $\rho'$-meson mass extracted in our previous paper on the longitudinally polarized $\rho$-meson DA [8], $m_{\rho'} = 1496 \pm 37$ MeV, rather close to the Particle Data Group value $m_{\rho'} = 1465 \pm 22$ MeV [23].

In the case of $b_1$-meson, one can analyze only the SR for the zeroth moment (decay constant $f^T_{b_1}$) of the DA (see Fig.3), the SRs for higher moments appearing to be invalid.

## 5 Processing different SRs and comparison of the results

We start with considering the results of processing both the types of SRs for $f^T_\rho$. Its dependence on the Borel parameter $M^2$ obtained from the "mixed parity" NLC SR, Eq. (15), with $s_0 = 2.9$ GeV$^2$ is shown in Fig. 1(a). Figure 1(b) shows $f^T_\rho$ as a function of the Borel parameter $M^2$ obtained from the "pure parity" NLC SR, Eq. (16), with $s_0 = 2.8$ GeV$^2$. Both kinds of SRs are rather sensitive to the $\rho'$-meson contribution and, for this reason, they were processed with taking it into account (see numerical results in Table 1). Solid lines correspond to the optimal thresholds $s_0$; the dashed lines — to the curves with the 10-fold variation of $\chi^2_{\min}$ (this corresponds approximately to the 5%-variation of $s_0$; definition of $\chi^2$, see in Appendix C, Eq.(C.1)). So, one can conclude that both types of NLC SRs agree rather well about the value of $f_\rho^T$. Note that the presented $f_\rho^T$ is rather close to the standard estimation $f_\rho^T = 0.160(10)$ GeV [4] and to the lattice one $f_{\rho Latt}^T(4\text{GeV}^2) = 0.165(11)$ GeV [24], and differs significantly from the result $f_\rho^T = 0.140$ GeV in [25].

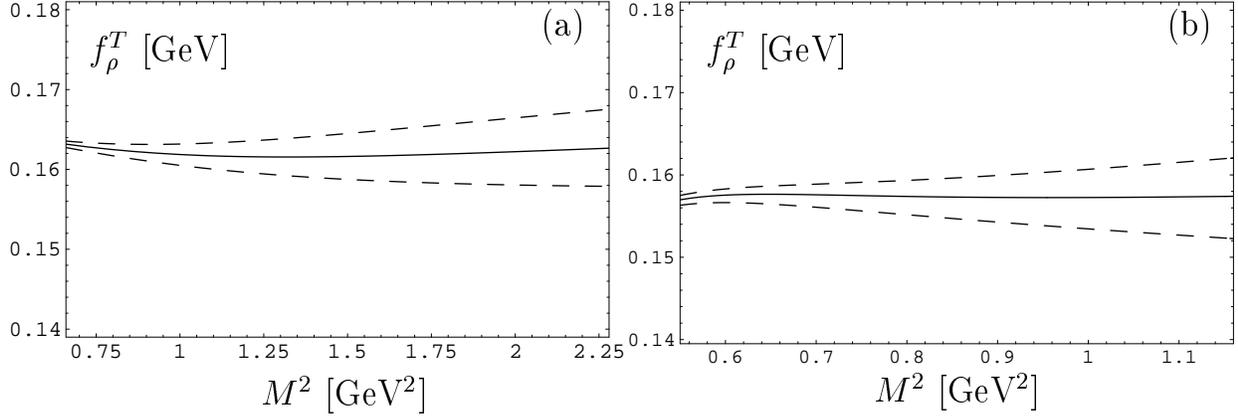

Figure 1: $f_\rho^T$ as a function of the Borel parameter $M^2$ obtained from: (a) the "mixed parity" NLC SR, Eq. (15), with $s_0 = 2.9$ GeV$^2$; (b) the "pure parity" NLC SR, Eq. (16), with $s_0 = 2.8$ GeV$^2$. The fidelity windows for both figures coincide with the whole depicted range of $M^2$.

| Type of SR | $f_M\left(1\text{GeV}^2\right)$ | $N=2$ | $N=4$ | $N=6$ | $N=8$ |
|---|---|---|---|---|---|
| | **Table 1**: The moments $\langle \xi^N \rangle_M(\mu^2)$ at $\mu^2 \sim 1$ GeV$^2$ (errors are depicted in brackets in a standard manner) | | | | |
| Asympt. WF | 1 | 0.2 | 0.086 | 0.047 | 0.030 |
| NLC SR Eq.(16) : $\rho^T$ | 0.157(5) | 0.296(20) | 0.196(6) | 0.132(5) | 0.089(4) |
| NLC SR Eq.(15) : $\rho^T$ | 0.162(5) | 0.329(11) | – | – | – |
| B&B SR : $\rho^T$ | 0.160(10) | 0.304(40)[4] | does not work | | |
| NLC SR Eq.(16) : $\rho'^T$ | 0.140(10) | 0.086(6) | 0.010(1) | 0.013(1) | 0.022(2) |
| NLC SR Eq.(17) : $b_1^T$ | 0.184(5) | does not work | | | |
| NLC SR Eq.(15) : $b_1^T$ | 0.181(5) | 0.144(15) | – | – | – |
| B&B SR : $b_1^T$ | 0.175(5) | does not work | | | |

Now we consider the results of processing SRs for the second moment $\langle \xi^2 \rangle_\rho^T$. First, we demonstrate the results of the "standard" approach: $\langle \xi^2 \rangle_\rho^T$ from Eq.(3.21) in [4] as a function of $M^2$ is shown in Fig.2(b) by a long-dashed line. This curve is not stable in $M^2$ at all, therefore the SR can provide merely a range of admissible values, $0.27 \leq \langle \xi^2 \rangle_\rho^T \leq 0.4$. As it is evident from Fig.2, this wide window agrees reasonably with both the estimates from the "mixed" (a) and "pure" (b) NLC SRs.

---

[4]The estimate presented in this cell has been obtained by processing the "mixed parity" SR established in [4], whereas in the original paper [4] this value amounts to 0.27(4).

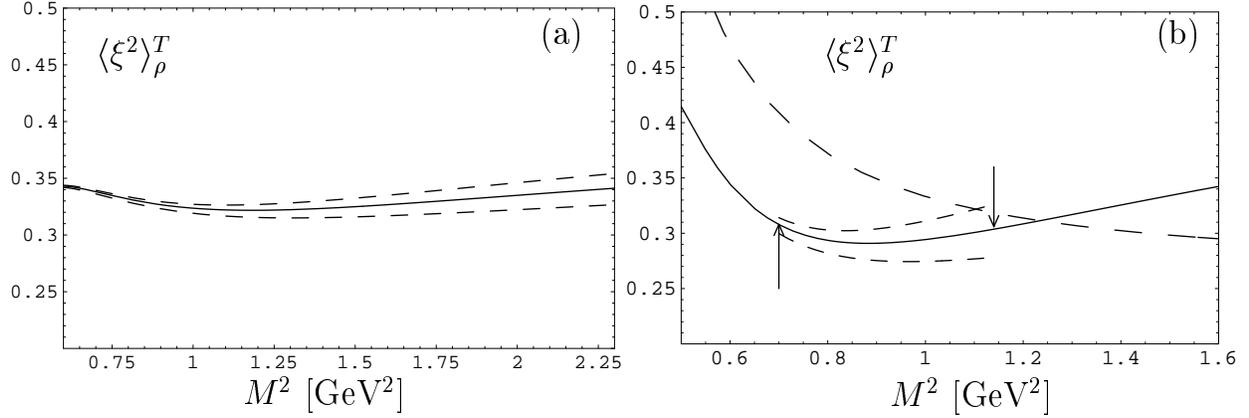

Figure 2: $\langle \xi^2 \rangle_\rho^T$ as a function of the Borel parameter $M^2$ obtained from: (a) the "mixed parity" NLC SR, Eq. (15), with $s_0 = 2.9$ GeV$^2$; (b) the "pure parity" NLC SR, Eq. (16), with $s_0 = 2.8$ GeV$^2$. Both kinds of SRs were processed with taking the $\rho'$-meson into account, the arrows show the fidelity window (for figure (a) the window coincides with the whole depicted range of $M^2$). Solid lines correspond to the optimal thresholds $s_0$, the short-dashed lines on both the figures correspond to the curves with the 10%-variation of $s_0$ (a) or of $\chi^2_{\min}$ (b). The long-dashed line in figure (b) represents the SR of Ball–Braun [4].

Note, the authors of [4] dealt with the quantity $a_2$, the Gegenbauer coefficient in the expansion of DA. The second moment of DA is trivially connected with this coefficient, $\langle \xi^2 \rangle = 0.2 + (12/35)a_2$. Using the SR of [4] for $a_2$, we obtain the corresponding window, $0.2 \leq a_2 \leq 0.4$, that leads to the mean value $\langle \xi^2 \rangle_{\rho \; [\text{Stand}]}^T = 0.30$ being surprisingly close to our estimate from NLC SRs, (see Table 1). However Ball and Braun have obtained the erroneous estimate $a_2 = 0.2 \pm 0.1$ producing, instead, the mean value $\langle \xi^2 \rangle_{\rho \; [\text{B\&B}]}^T = 0.27$.

The curves for the next higher moments, whose estimates are presented in Table 1, have the fidelity windows and the stability behavior similar to $\langle \xi^2 \rangle_\rho^T(M^2)$ in Fig.2(b). Finally, in Fig.3, we demonstrate a very good correspondence between the values of $f_{b_1}^T$ obtained in different NLC SRs.

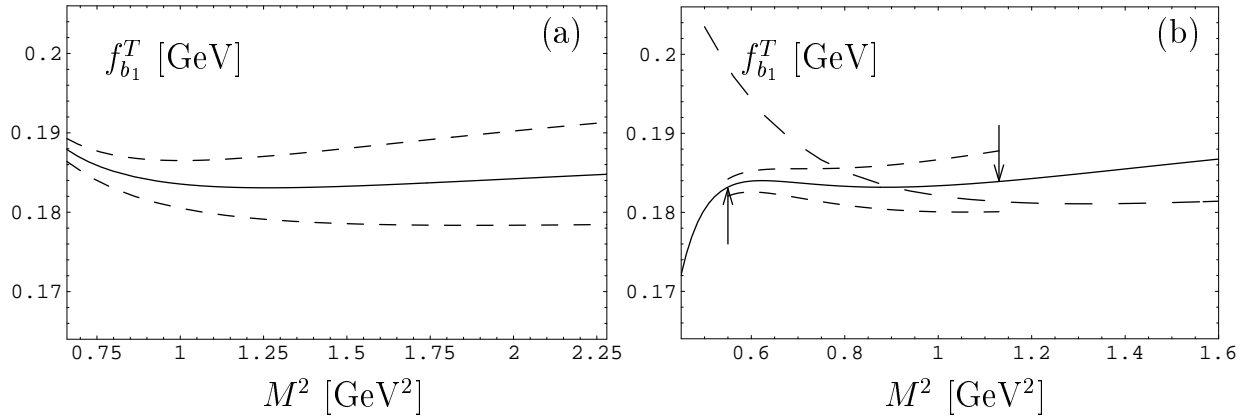

Figure 3: The curves $f_{b_1}^T$ in $M^2$ obtained from: (a) the "mixed parity" NLC SR (with taking the $\rho'$-meson into account with $f_{\rho'}$ defined from "pure parity" SR (16)); (b) the "pure parity" NLC SR (17). The arrows show the fidelity window (for the right figure, the window coincides with the whole depicted range of $M^2$). Solid lines correspond to the optimal thresholds; the short-dashed lines on both the figures, – to the curves with the 10-fold variation of $\chi^2_{\min}$; the long-dashed line on the right figure corresponds to the real B&B curve.

# 6 DA models and their check

Possible models of DAs corresponding to the moments in Table 1 are of the form

$$\varphi_\rho^{T,mod}(x,\mu^2) = 1.382\,[\varphi^{as}(x)]^2\left(1+0.927 C_2^{3/2}(\xi)+0.729 C_4^{3/2}(\xi)\right)$$
$$= \varphi^{as}(x)\left(1+0.29 C_2^{3/2}(\xi)+0.41 C_4^{3/2}(\xi)-0.32 C_6^{3/2}(\xi)\right), \quad (21)$$
$$\varphi_{\rho'}^{T,mod}(x,\mu^2) = \varphi^{as}(x)\left(1-0.339 C_2^{3/2}(\xi)+0.003 C_4^{3/2}(\xi)+0.192 C_6^{3/2}(\xi)\right), \quad (22)$$
$$\varphi_{b_1}^{mod}(x,\mu^2) = \varphi^{as}(x)\left(1-(0.175\pm 0.05) C_2^{3/2}(\xi)\right), \quad (23)$$

where $\xi \equiv 1-2x$, $C_n^\nu(\xi)$ are the Gegenbauer polynomials (GP), and the norm $\mu^2 \simeq 1$ GeV$^2$ corresponds to a mean value of $M^2$. Recall again that the value of the important coefficient $a_2 = 0.29$ in (21) is confirmed by 3 sources: "pure" NLC SR (16), "mixed" NLC SR (15), and a mean value from the "mixed" standard SR. Figures 4, 5(a) contain curves of DA corresponding to $\rho_\perp$, eqs. (21), and $\rho'_\perp$ (22). The arising 3-hump shape of DA for $\rho_\perp$ drastically differs from that obtained in [4] and from the one obtained in chiral effective theory [25].

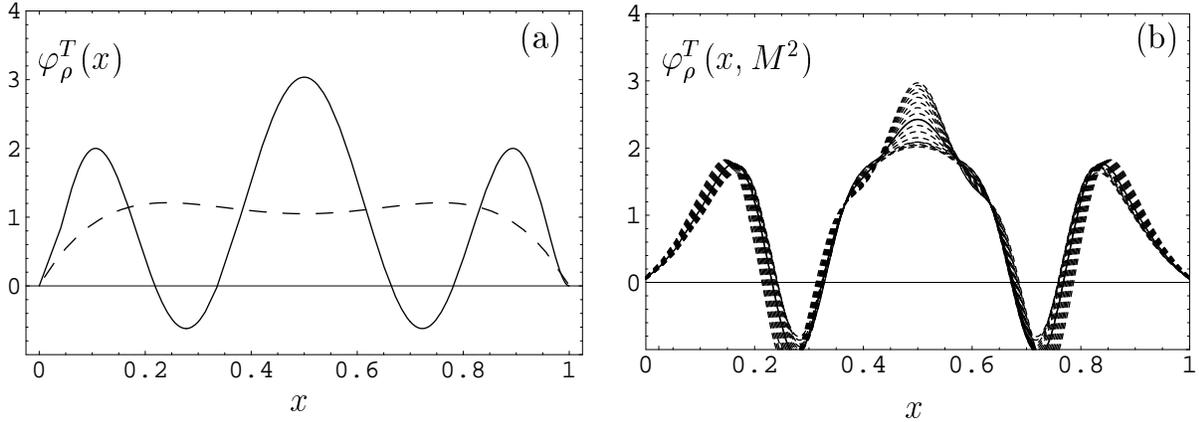

Figure 4: (a) The curves of $\varphi_\rho^{T,mod}(x,1\text{ GeV}^2)$: Solid lines correspond to the best fits for determined moments (see Table 1); the dashed line on the left figure corresponds to the B&B curve (which fits only $\langle \xi^2\rangle_\rho^T \approx 0.27$). (b) The rhs of Eq.(16) SR$_\rho^T(x,M^2)$ in $x$. Different lines here correspond to different values of Borel $M^2 = 0.7 - 0.9$ GeV$^2$.

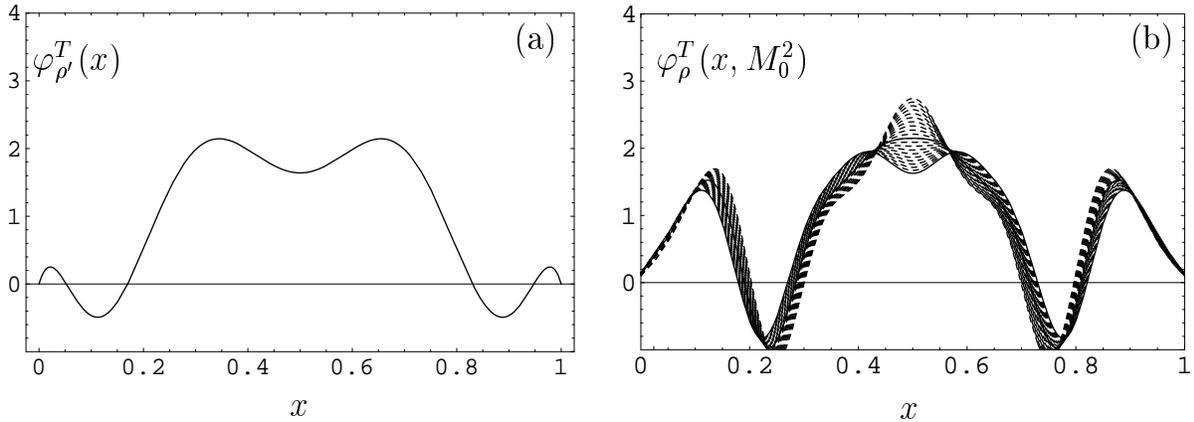

Figure 5: (a) The curve of $\varphi_{\rho'}^{T,mod}(x,1\text{ GeV}^2)$ in $x$. (b) The rhs of Eq.(16) SR$_\rho^T(x,M_0^2)$ in $x$. Solid and dashed lines here correspond to different values of nonlocality parameter $\lambda_q^2 = 0.4 - 0.5$ GeV$^2$ with fixed value of Borel parameter $M_0^2 = 0.8$ GeV$^2$.

This difference mainly appears due to the higher moments, $N = 4, 6, 8$, involved into consideration. Nevertheless, the hump shape is not an artifact of the GP expansion series truncation. These models really contain only 3 first GPs, meanwhile, it is enough to reproduce all 4 moments up to $N = 8$. Moreover, an additionally smoothed[5] rhs of the NLC SR (16) demonstrates qualitatively the same behaviour in $x$ (at admissible $M^2$) as the model DA, compare Figs. 4(a) and (b). The stability of the DA shape with respect to the variation of ansatz is also checked. To this end, we show in Fig. 5(b) the same r.h.s. of (16) as in Fig. 4(b), but with different values of the single ansatz parameter $\lambda_q^2 = 0.4 - 0.5$ GeV$^2$ at fixed value $M_0^2 = 0.8$ GeV$^2$.

Inverse moments of DAs often appear in perturbative QCD predictions for exclusive reactions. The estimates for important $\langle x^{-1} \rangle_M$ moments obtained from the model DAs are presented here[6]

$$\langle x^{-1} \rangle_\rho \equiv \int_0^1 \frac{\varphi_\rho^T(x, 1 \text{ GeV}^2)}{x} \, dx \;\; = \;\; \begin{cases} 4.15^{+0.4}_{-0.1} & \text{(here)} \\ 3.6 & \text{(B\&B model)} \end{cases} \quad (24)$$

$$\langle x^{-1} \rangle_{\rho'} \equiv \int_0^1 \frac{\varphi_{\rho'}^T(x, 1 \text{ GeV}^2)}{x} \, dx \;\; = \;\; 2.57 \pm 0.20 \text{ (here)} \quad (25)$$

$$\langle x^{-1} \rangle_{b_1} \equiv \int_0^1 \frac{\varphi_{b_1}^T(x, 1 \text{ GeV}^2)}{x} \, dx \;\; = \;\; 2.48 \pm 0.20 \text{ (here)} \quad (26)$$

It is useful to construct **an independent** SR for these inverse moments to verify the DA models (21, 22, 23). Namely, the weighted sum $C(M^2)$ of these moments

$$C(M^2) \equiv \langle x^{-1} \rangle_\rho + \langle x^{-1} \rangle_{\rho'} \left(\frac{f_{\rho'}^T}{f_\rho^T}\right)^2 e^{-(m_{\rho'}^2 - m_\rho^2)/M^2} + \langle x^{-1} \rangle_{b_1} \left(\frac{f_{b_1}^T}{f_\rho^T}\right)^2 e^{-(m_{b_1}^2 - m_\rho^2)/M^2} \quad (27)$$

can be obtained by integrating the rhs of the "mixed" NLC SR (15) with the weight $1/x$.

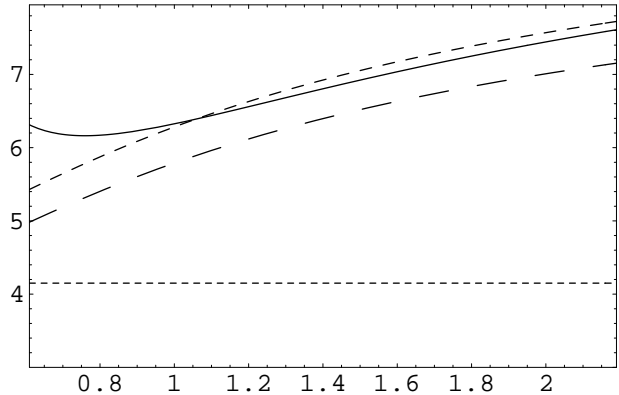

Figure 6: $C(M^2)$ as a function of $M^2$ (solid line) determined by Eq.(27) and integrating in $x$ Eq.(15) in comparison with the lhs of Eq.(28) (long-dashed line). The dotted line corresponds to $\langle x^{-1} \rangle_\rho = 4.15$; whereas the dashed line – to the lhs of Eq.(28) with upper values of corresponding moments.

A comparison of the function $C(M^2)$ with the corresponding combination of model estimates (24, 25, 26) obtained in different kinds of NLC SRs (mainly from the "pure" ones) leads to an approximate equation

$$4.15 + 2.57 \left(\frac{f_{\rho'}^T}{f_\rho^T}\right)^2 e^{-(m_{\rho'}^2 - m_\rho^2)/M^2} + 2.48 \left(\frac{f_{b_1}^T}{f_\rho^T}\right)^2 e^{-(m_{b_1}^2 - m_\rho^2)/M^2} \approx C(M^2) \,, \quad (28)$$

illustrated in Fig. 6.

As a result, one can conclude:

---

[5]A certain smoothing of some $\delta$-functions in the r.h.s. of the SR (see Appendix B) is not important.
[6]The upper error +0.4 in (24) corresponds to an overestimate $\langle \xi^2 \rangle = 0.329$ from the "mixed" SR

1. The "mixed" NLC SR is highly sensitive to $b_1$- and $\rho'$-meson contributions, the difference in the behavior of $C(M^2)$ (solid line) and in the $\rho$-contribution alone (dotted line) illustrates this point.

2. The curve $C(M^2)$ lies between mean and upper estimates for the lhs of (28), so it is in reasonable agreement with the estimates (24, 25, 26). It also demonstrates an overestimation of DA moments in the "mixed" SR as compared to that obtained from the "pure" one.

# 7 DA models and the $B \to \rho e\nu$ decay form factors

The new DA shapes result in different pQCD predictions for exclusive reactions with the $\rho$-meson. As an example, we re-estimate form factors $V(t)$, $A_{1,2}(t)$ corresponding to the transition matrix element $\langle \rho, \lambda|(V-A)_\mu|B\rangle$ of the process $B \to \rho \, e\nu$, in the framework of the light-cone SR approach, [26]. That was done earlier by Ball and Braun in [27], [28] on the base of DAs from [4]. Thus, to estimate the influence of the new nonperturbative input presented in the previous sections, we have used the LC SR in the leading twist approximation (cf. [27]). Just as in the case of the LC expansion for the transition amplitude $\gamma^*\gamma \to \pi^0$, one might expect high sensitivity to the end-point behavior of the DAs, as they enter into convolution integrals like $\langle x^{-1}\rangle_M$ estimated in (24).

However, there are some essential differences which effectively soften our expectations. First, the DAs also enter into the "phenomenological" side of the SR in the "continuum" contribution of higher excited states in the channel with $B$-meson quantum numbers. This, actually, is a specific feature of any LC SR. By subtracting the "continuum", one actually obtains "infrared safe quantities" like $\int_\epsilon^1 dx \varphi(x)/x$ where $\epsilon \simeq (m_b^2 - t)/(s_0^B - t)$, $m_b \simeq 4.8\,\text{GeV}$, and $s_0^B \simeq 34$ GeV$^2$ is the continuum threshold in the $B$-channel[7] as defined from the 2-point QCD SRs for the $B$-meson decay constant $f_B$ (see [29]). For $t \approx 0$, $\epsilon \simeq 0.5-0.6$ and the LC SR should not be so sensitive to the end-point region $x \sim 0$. Obviously, the end-point region becomes to be important for higher momentum transfers $t$. However, for $t \geq 20$ GeV$^2$ the LC expansion would hardly make sense. The second factor which eventually decreases the importance of the end-point region is connected with the standard Borel transformation of the SR with respect to the virtuality of the $B$-meson current: $-p_B^2 \to M_B^2$. The corresponding contribution from the coefficient function produces a standard suppression exponent: $\exp(\bar{x}(t - m_b^2)/x M_B^2)$. Numerically, it occurred to be less important.

We have treated the LC SRs using the same input parameters and the same procedure of extracting the physical form factors as in Ref.[27]. However, if one tries to fix the onset of the "continuum" by hand to the value $s_0^B \simeq 34$ GeV$^2$ dictated by the 2-point SRs for $f_B$, one encounters **inadmissible** uncertainties in the determination of the form factors when using our new nonperturbative input DAs. To get a stable SR, one is forced to allow a higher value for the $s_0^B$ parameter.

Below, the form factor values are written at a zero momentum transfer ($t=0$) as compared with B&B results:

$$V(0) = \begin{cases} 0.37(1) & (\text{here } [s_0 = 50 \text{ GeV}^2],\ \chi^2 \approx 0.4) \\ 0.35(2) & ([27]\ [s_0 = 34 \text{ GeV}^2],\ \chi^2 \approx 3.4) \end{cases}$$
$$A_1(0) = \begin{cases} 0.283(4) & (\text{here } [s_0 = 45 \text{ GeV}^2],\ \chi^2 \approx 0.1) \\ 0.27(1) & ([27]\ [s_0 = 34 \text{ GeV}^2],\ \chi^2 \approx 1.1) \end{cases} \quad (29)$$
$$A_2(0) = \begin{cases} 0.30(1) & (\text{here } [s_0 = 50 \text{ GeV}^2],\ \chi^2 \approx 0.2) \\ 0.28(1) & ([27]\ [s_0 = 34 \text{ GeV}^2],\ \chi^2 \approx 1.1) \end{cases}$$

Our form factors are slightly higher than those in [27] and possess a better accuracy (compare $\chi^2$ in (30)). The difference becomes more pronounced for a large value of the momentum transfer $t$, $(m_b^2 - t \sim \mathcal{O}(m_b))$. The last is not surprising due to higher sensitivity to the end-point behavior of the input DA in this region. The form factors presented are determined with new "optimal" thresholds $s_0^B$ providing few times better processing accuracy. Note that the parameters of the usual "pole" parameterization of the form factors change significantly as compared to that in [27], *e.g.*,

$$A_1(t) = \frac{0.283}{1 - 0.157(t/m_B^2) - 0.837(t/m_B^2)^2}$$

---

[7] As we shall see below, the LC SRs "prefer" a higher value.

The important form factor $A_1(t)$ (solid line) increases about $5 - 10\%$ in comparison with the B&B result (the bars in the figure show the errors of the B&B calculations), with an optimal threshold $s_0^B \simeq 45$ GeV$^2$.

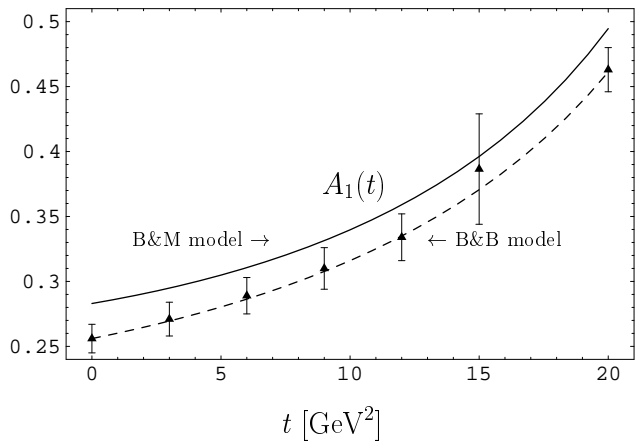

Figure 7: Form factor $A_1(t)$; solid line corresponds to our processing of LC QCD SR, dashed line – to processing following the B&B formulas ([27]) (the bars in the figure show the errors of the B&B calculations).

## 8  Conclusion

Let us summarize the main results of this paper:

1. We construct NLC SRs for DA for each P-parity channels, based on the properties of the duality transformation. The negative parity NLC SR for transversely polarized $\rho$-, $\rho'$-mesons works rather well and allows us to estimate the 2-nd, 4-th, 6-th, and 8-th moments of the leading twist DAs. The positive parity SR for the transversely polarized $b_1$-meson can provide only the value of the $b_1$-meson lepton decay constant, $f_{b_1}^T$. It should be emphasized that an analogous evaluation of the moments within the standard QCD SR approach is impossible.

2. Results of processing different NLC SRs of the "pure" (see Figs. 1b, 2b, 3b) and "mixed" (see Figs. 1a, 2a, 3a) parity are compared, and a reasonable agreement between them is found. The "mixed" SR in the standard version admits merely a window of possible values of the second moment $\langle \xi^2 \rangle$ (see, *e.g.* , [4]); the position of the window is corrected here and, as a result, agrees with the NLC SR results presented in Table 1.

3. The models for the leading twist DAs of the $\rho_\perp$- and $\rho'_\perp$-mesons, (21,22), and of the $b_1^\perp$-meson, (23), are suggested. The shape of a new $\rho_\perp$-meson distribution (see Fig. 4a) drastically differs from that obtained by Ball and Braun [4] only on the basis of the value $a_2 = 0.2$. The latter estimate is discussed in sect. 5.

4. We estimate important integrals appearing in perturbative QCD predictions for different exclusive reactions, $\langle x^{-1}\rangle_M \equiv \int_0^1 \frac{\varphi_M^T(x)}{x}\,dx$ in (24)-(26), based on our results for the DA shapes. We check the self-consistency of these results by comparing them with those obtained from an independent "mixed" QCD SR for the inverse moment $\langle x^{-1}\rangle_M$ and find an agreement.

5. Form factors of the process $B \to \rho\, e\nu$, $V(t)$, $A_{1,2}(t)$ where $t$ is momentum transfer are also re-estimated in the framework of the light-cone SR approach [27] on the basis of the new model for the $\rho$-meson DAs; the results are slightly higher and have uncertainties a few times as small as those obtained by Ball and Braun.

Finally, we can conclude that the nonlocal condensate QCD SR approach to distribution amplitudes is self-consistent and gives reliable results. An open problem of this approach is to determine well-established models of distribution functions $f_\Gamma(\nu)$ from the theory of nonperturbative QCD vacuum. First direct attempts to calculate quark NLC have been done in lattice simulations in [21]. The "short distance" correlation length of NLC has also been extracted later in [22]; it turns out to be reasonably close to the value of $1/\lambda_q$ and confirms the validity of our Gaussian NLC model.

### Acknowledgments

This work was partially supported by the Russian Foundation for Basic Research (grant N 00-02-16696) and Heisenberg–Landau Program. We are grateful to R. Ruskov for help in a better understanding of $B$ decay (sect. 8), to O. V. Teryaev, M. Polyakov, and N. G. Stefanis for fruitful discussions. We gratefully acknowledge the warm hospitality of Prof. K. Goeke and Dr. Dr. N. G. Stefanis at Bochum University, where this work was completed.

## Appendix

## A  Decomposition of rank-4 tensor $\Pi_{(N)}^{\mu\nu;\alpha\beta}$

$$P_1^{\mu\nu;\alpha\beta} \equiv \frac{1}{2q^2}\left[g^{\mu\alpha}q^\nu q^\beta - g^{\nu\alpha}q^\mu q^\beta - g^{\mu\beta}q^\nu q^\alpha + g^{\nu\beta}q^\mu q^\alpha\right]; \tag{A.1}$$

$$P_2^{\mu\nu;\alpha\beta} \equiv \frac{1}{2}\left[g^{\mu\alpha}g^{\nu\beta} - g^{\mu\beta}g^{\nu\alpha}\right] - P_1^{\mu\nu;\alpha\beta}; \tag{A.2}$$

$$Q_1^{\mu\nu;\alpha\beta} \equiv \frac{1}{2(qz)}\left[g^{\mu\alpha}q^\nu z^\beta + g^{\nu\beta}q^\mu z^\alpha - g^{\mu\beta}q^\nu z^\alpha - g^{\nu\alpha}q^\mu z^\beta\right]; \tag{A.3}$$

$$Q_3^{\mu\nu;\alpha\beta} \equiv \frac{1}{2(qz)}\left[g^{\mu\alpha}z^\nu q^\beta + g^{\nu\beta}z^\mu q^\alpha - g^{\mu\beta}z^\nu q^\alpha - g^{\nu\alpha}z^\mu q^\beta\right]; \tag{A.4}$$

$$Q_z^{\mu\nu;\alpha\beta} \equiv \frac{q^2}{2(qz)^2}\left[g^{\mu\alpha}z^\nu z^\beta + g^{\nu\beta}z^\mu z^\alpha - g^{\mu\beta}z^\nu z^\alpha - g^{\nu\alpha}z^\mu z^\beta\right]; \tag{A.5}$$

$$Q_q^{\mu\nu;\alpha\beta} \equiv \frac{1}{2(qz)^2}\left(q^\alpha z^\beta - q^\beta z^\alpha\right)\left(q^\mu z^\nu - q^\nu z^\mu\right). \tag{A.6}$$

$$g^{\mu\alpha}z^\nu z^\beta P_1^{\mu\nu;\alpha\beta} \equiv P_1^{\mu z;\mu z} = -P_2^{\mu z;\mu z} = \frac{(qz)^2}{q^2};\ Q_1^{\mu z;\mu z} = Q_3^{\mu z;\mu z} = Q_z^{\mu z;\mu z} = Q_q^{\mu z;\mu z} = 0; \tag{A.7}$$

$$q^\mu q^\alpha z^\nu z^\beta P_1^{\mu\nu;\alpha\beta} \equiv P_1^{qz;qz} = Q_1^{qz;qz} = Q_3^{qz;qz} = -Q_q^{qz;qz} = -\frac{(qz)^2}{2};\ P_2^{qz;qz} = Q_z^{qz;qz} = 0. \tag{A.8}$$

Let us write down the parameterization of matrix elements of a composite tensor current operator, see, *e.g.* , [28]:

$$\langle 0 \mid \bar{d}(z)\sigma_{\mu\nu}u(0) \mid \rho_\perp(p,\lambda)\rangle\Big|_{z^2=0} = if_{\rho_\perp}^T \left[ (\varepsilon_\mu(p,\lambda)p_\nu - \varepsilon_\nu(p,\lambda)p_\mu) \int_0^1 dx\, \varphi_\rho^T(x)\, e^{ix(zp)} \right.$$

$$+ (\varepsilon_\mu(p,\lambda)z_\nu - \varepsilon_\nu(p,\lambda)z_\mu) p^2 \int_0^1 dx\, V_1(x)\, e^{ix(zp)}$$

$$\left. + (p^\mu z_\nu - p^\nu z_\mu)(\varepsilon(p,\lambda)z)p^2 \int_0^1 dx\, V_2(x)\, e^{ix(zp)} \right] \quad \text{(A.9)}$$

$$\langle 0 \mid \bar{d}(z)\sigma_{\mu\nu}u(0) \mid b_1(p,\lambda)\rangle\Big|_{z^2=0} = f_{b_1}^T \left[ \epsilon_{\mu\nu\alpha\beta}\varepsilon^\alpha(p,\lambda)p^\beta \int_0^1 dx\, \varphi_{b_1}(x)\, e^{ix(zp)} \right.$$

$$+ \epsilon_{\mu\nu\alpha\beta}\varepsilon^\alpha(p,\lambda)z^\beta p^2 \int_0^1 dx\, U_1(x)\, e^{ix(zp)}$$

$$\left. + \epsilon_{\mu\nu\alpha\beta}p^\alpha z^\beta (\varepsilon(p,\lambda)z)p^2 \int_0^1 dx\, U_2(x)\, e^{ix(zp)} \right] . \quad \text{(A.10)}$$

Here we decode our shorthand notation used in Section 2:

$$v_0 \equiv \left|f_{\rho_\perp}^T\right|^2 \langle x^N \rangle_{\rho_\perp} ; \qquad v_1 \equiv \left|f_{\rho_\perp}^T\right|^2 \langle -iNx^{N-1}\rangle_{V_1} ; \qquad v_2 \equiv \left|f_{\rho_\perp}^T\right|^2 \langle -N(N-1)x^{N-2}\rangle_{V_2} ;$$

$$u_0 \equiv \left|f_{b_1}^T\right|^2 \langle x^N \rangle_{b_1} ; \qquad u_1 \equiv \left|f_{b_1}^T\right|^2 \langle -iNx^{N-1}\rangle_{U_1} ; \qquad u_2 \equiv \left|f_{b_1}^T\right|^2 \langle -N(N-1)x^{N-2}\rangle_{U_2} ,$$

(with $\langle f(x)\rangle_U \equiv \int_0^1 dx\, f(x)U(x)$). In the general case, the whole system of equations for different twist DA contributions is of the following form

$$\frac{\Pi_-(q^2,qz)}{2(qz)^N q^2} = -v_0 + u_1 + u_2 ; \quad \frac{K_1(q^2,qz)}{2(qz)^N q^2} = -v_1 - u_2 ; \quad \frac{K_z(q^2,qz)}{2(qz)^N q^2} = +u_2 ; \quad \text{(A.11)}$$

$$\frac{\Pi_+(q^2,qz)}{2(qz)^N q^2} = +u_0 + u_1 + u_2 ; \quad \frac{K_3(q^2,qz)}{2(qz)^N q^2} = -u_1 - u_2 ; \quad \frac{K_q(q^2,qz)}{2(qz)^N q^2} = v_2 - u_2 . \quad \text{(A.12)}$$

# B Expressions for nonlocal contributions to SR

To construct SR for distribution amplitudes, it is useful to parameterize NLC behaviors by the "distribution functions" [7, 8, 13, 14] *a'la* $\alpha$-representation of propagators, *e.g.* , $f_S(\alpha)$ for the scalar condensate $M_S(z^2)$ [8]

$$M_S\left(z^2\right) = \langle \bar{q}(0)q(0)\rangle \int_0^\infty e^{\alpha z^2/4} f_S(\alpha)\, d\alpha, \text{ where } \int_0^\infty f_S(\alpha)\, d\alpha = 1,\ \int_0^\infty \alpha f_S(\alpha)d\alpha = \frac{\lambda_q^2}{2}, \quad \text{(B.1)}$$

and for the vector condensate $M_V^\mu(z^2)$,

$$M_V^\mu(z) \equiv \langle \bar{q}(0)\gamma^\mu q(z)\rangle = -iz^\mu \frac{A_S}{4} \int_0^\infty e^{\alpha z^2/4} f_V(\alpha)\, d\alpha, \text{ where } \int_0^\infty f_V(\alpha)\, d\alpha = 0. \quad \text{(B.2)}$$

Here and in the following we take quark and gluon fields in the fixed point gauge $z^\mu A_\mu(z) = 0$ where the path-ordered exponential $E(0,z) = 1$. The appearing in the SR quark-gluon-quark NLC $M_{\mu\nu}(\tilde{M}_{\mu\nu})$,

$$M_{\mu\nu}(y,z) \equiv \langle \bar{q}(0)\gamma_\nu \hat{A}_\mu(z)q(y)\rangle = \left(y_\mu z_\nu - g_{\mu\nu}(zy)\right) \cdot M_{T1} + \left(z_\mu z_\nu - g_{\mu\nu}z^2\right) \cdot M_{T2} + \cdots \text{(B.3)}$$

$$\tilde{M}_{\mu\nu}(y,z) \equiv \langle \bar{q}(0)\gamma_\nu(\gamma_5)\hat{A}_\mu(z)q(y)\rangle = \varepsilon_{\mu\nu\rho\sigma}z_\rho y_\sigma \cdot M_{T3} + \cdots, \quad \text{(B.4)}$$

---

[8] In deriving these sum rules we can always make a Wick rotation, i.e., we assume that all coordinates are Euclidean, $z^2 < 0$.

can be decomposed in form factors $M_{T1-T3}$, where the tensors in front of them satisfy the gauge condition $z^\mu M_{\mu\nu}(\tilde{M}_{\mu\nu}) = 0$ ( since $z^\mu \hat{A}_\mu(z) = 0$). The NLC $M_{T1-T3}$ can be parameterized by a triple integral representation

$$M_{Ti}(z^2, y^2, (z-y)^2) = A_{Ti} \int_0^\infty e^{(\alpha_1 z^2/4 + \alpha_2 y^2/4 + \alpha_3 (z-y)^2/4)} f_i(\alpha_1, \alpha_2, \alpha_3)\, d\alpha_1 d\alpha_2 d\alpha_3, \qquad (B.5)$$

where $A_{Ti} = \{-\frac{3}{8}A_S, \frac{1}{2}A_S, \frac{3}{8}A_S\}$, and $A_S = \frac{8\pi}{81} \langle \sqrt{\alpha_s} \bar{q}(0) q(0) \rangle^2$. The function $f_S(\alpha)$ and other similar functions $f_\Gamma(\alpha)$ describe distributions of vacuum fields in virtuality $\alpha$ for every type ($\Gamma$) of NLC. The convolutions $\Delta\Phi_\Gamma(x, M^2)$ of the distribution functions $f_\Gamma$ and coefficient functions completely determine the r.h.s. of SR's, so $\Delta\Phi_\Gamma$ depends on the model of $f_\Gamma$. For vacuum distribution functions $f_\Gamma(\alpha)$, we use the set of the simplest ansatzes

$$f_S(\alpha) = \delta\left(\alpha - \lambda_q^2/2\right); \qquad f_V(\alpha) = \delta'\left(\alpha - \lambda_q^2/2\right); \qquad (B.6)$$

$$f_{T_{1,2,3}}(\alpha_1, \alpha_2, \alpha_3) = \delta\left(\alpha_1 - \lambda_q^2/2\right) \delta\left(\alpha_2 - \lambda_q^2/2\right) \delta\left(\alpha_3 - \lambda_q^2/2\right). \qquad (B.7)$$

Their meaning and relation to initial NLCs have been discussed in detail in [6, 7]. The contributions $\Delta\Phi_\Gamma(x, M^2)$ to the r.h.s of SR, corresponding to these ansatzes, are shown below. The limit of these expressions to the standard (local) contributions $\varphi_\Gamma(x, M^2)$ – $\lambda_q^2 \to 0$, $\Delta\Phi_\Gamma(x, M^2) \to \Delta\varphi_\Gamma(x, M^2)$ are also written for comparison. Hereafter $\Delta \equiv \lambda_q^2/(2M^2)$, $\bar{\Delta} \equiv 1 - \Delta$:

$$\Delta\Phi_S\left(x, M^2\right) = \frac{A_S}{M^4} \frac{18}{\bar{\Delta}\Delta^2} \{\theta(\bar{x} > \Delta > x)\, \bar{x}\, [x + (\Delta - x)\ln(\bar{x})] + (\bar{x} \to x) +$$
$$+ \theta(1 > \Delta)\theta(\Delta > x > \bar{\Delta})\, [\bar{\Delta} + (\Delta - 2\bar{x}x)\ln(\Delta)]\}, \qquad (B.8)$$

$$\Delta\varphi_S\left(x, M^2\right) = \frac{A_S}{M^4} 9\, (\delta(x) + (\bar{x} \to x));$$

$$\Delta\Phi_V\left(x, M^2\right) = \frac{A_S}{M^4} (x\delta'(\bar{x} - \Delta) + (\bar{x} \to x)), \qquad (B.9)$$

$$\Delta\varphi_V\left(x, M^2\right) = \frac{A_S}{M^4} (x\delta'(\bar{x}) + (\bar{x} \to x)); \qquad (B.10)$$

$$\Delta\Phi_{T_1}\left(x, M^2\right) = -\frac{3A_S}{M^4}\theta(1 > 2\Delta) \left\{ [\delta(x - 2\Delta) - \delta(x - \Delta)]\left(\frac{1}{\Delta} - 2\right) + \theta(2\Delta > x)\cdot \right.$$
$$\left. \theta(x > \Delta)\frac{\bar{x}}{\bar{\Delta}}\left[\frac{x - 2\Delta}{\Delta\bar{\Delta}}\right] \right\} + (\bar{x} \to x), \qquad (B.11)$$

$$\Delta\varphi_{T_1}\left(x, M^2\right) = \frac{3A_S}{M^4} (\delta'(\bar{x}) + (\bar{x} \to x));$$

$$\Delta\Phi_{T_2}\left(x, M^2\right) = \frac{4A_S}{M^4}\bar{x}\theta(1 > 2\Delta) \left\{ \frac{\delta(x - 2\Delta)}{\Delta} - \theta(2\Delta > x)\theta(x > \Delta)\cdot \right.$$
$$\left. \frac{1 + 2x - 4\Delta}{\bar{\Delta}\Delta^2} \right\} + (\bar{x} \to x), \qquad (B.12)$$

$$\Delta\varphi_{T_2}\left(x, M^2\right) = -\frac{2A_S}{M^4} (x\delta'(\bar{x}) + (\bar{x} \to x));$$

$$\Delta\Phi_{T_3}\left(x, M^2\right) = \frac{3A_S \bar{x}}{M^4 \bar{\Delta}\Delta} \left\{ \theta(2\Delta > x)\theta(x > \Delta)\theta(1 > 2\Delta)\left[2 - \frac{\bar{x}}{\Delta} - \frac{\Delta}{\bar{\Delta}}\right] \right\}$$
$$+ (\bar{x} \to x), \qquad (B.13)$$

$$\Delta\varphi_{T_3}\left(x, M^2\right) = \frac{3A_S}{M^4} (\delta(\bar{x}) + (\bar{x} \to x));$$

$$\Delta\Phi_G\left(x, M^2\right) = \frac{\langle \alpha_s GG \rangle}{24\pi M^2} (\delta(x - \Delta) + (\bar{x} \to x)), \qquad (B.14)$$

$$\begin{aligned}
\Delta\varphi_G\left(x,M^2\right) &= \frac{\langle\alpha_s GG\rangle}{24\pi M^2}\left(\delta\left(\bar{x}\right)+(\bar{x}\to x)\right); \\
\Delta\Phi'_G\left(x,M^2\right) &= \frac{\langle\alpha_s GG\rangle}{6\pi M^2}\frac{\theta\left(\Delta<x\right)\theta\left(x<1-\Delta\right)}{1-2\Delta}; \\
\Delta\varphi'_G\left(x,M^2\right) &= \frac{\langle\alpha_s GG\rangle}{6\pi M^2}.
\end{aligned} \qquad (B.15)$$

For quark and gluon condensates, we use the standard estimates $\langle\sqrt{\alpha_s}\bar{q}(0)q(0)\rangle \approx (-0.238 \text{ GeV})^3$, $\frac{\langle\alpha_s GG\rangle}{12\pi} \approx 0.001 \text{ GeV}^4$ [30] and $\lambda_q^2 = \frac{\langle\bar{q}(ig\sigma_{\mu\nu}G^{\mu\nu})q\rangle}{2\langle\bar{q}q\rangle} = 0.4\pm 0.1 \text{ GeV}^2$ normalized at $\mu^2 \approx 1 \text{ GeV}^2$.

**Expressions for perturbative spectral density**: Radiative corrections reach 10 % of the Born result at $s \sim 1 \text{ GeV}^2$.

$$\rho_T^{\text{pert}}(x,s) = \frac{3}{2\pi^2}x\bar{x}\left\{1+\frac{\alpha_s(\mu^2)C_F}{4\pi}\left(2\ln\left[\frac{s}{\mu^2}\right]+6-\frac{\pi^2}{3}+\ln^2(\bar{x}/x)+\ln(x\bar{x})\right)\right\}. \qquad (B.16)$$

Here $\mu^2 \sim 1 \text{ GeV}^2$ corresponds to the average value of the Borel parameter $M^2$ in the stability window; $\alpha_s\left(1\text{GeV}^2\right) \approx 0.52$. We also use the 'mixed' perturbative spectral density suggested in [31] in the "mixed" SR:

$$\rho_T^{\text{mixed}}(x,s;s_\rho^T,s_b^T) \equiv \rho_T^{\text{pert}}(x;s)\frac{1}{2}\left[\theta\left(s_\rho^T-s\right)+\theta\left(s_b^T-s\right)\right]. \qquad (B.17)$$

# C  About $\chi^2$-definition in Sum Rules

Let us discuss the definition of $\chi^2$ for the SR case. We have here the function $F(M^2,s)$, and the problem is to find the best value $s_0$, such that $F(M^2,s_0)$ is the most close to a constant value for $M_-^2 \leq M^2 \leq M_+^2$ (values of $M_\pm^2$ are known and fixed from standard constraints of QCD SR, see [30, 8] ). We define the function $\chi^2(s)$ for the curve $F(M^2,s)$ with $M^2 \in [M_-^2, M_+^2]$ in the following manner:

$$\chi^2(s) \equiv \frac{1}{(N-1)\epsilon^2}\sum_{k=0}^{N}\left[F\left(M_-^2+k\delta,s\right)-\frac{1}{N+1}\sum_{k=0}^{N}F\left(M_-^2+k\delta,s\right)\right]^2, \qquad (C.1)$$

where $\delta = (M_+^2-M_-^2)/N$, $N \simeq 10$, and $\epsilon$ is of an order of the last decimal digit in $F(M^2,s)$ we are interested in (in the case of decay constant $f_\rho \approx 200$ MeV, $\epsilon \approx 1$ MeV; in the case of the second moment $\langle\xi^2\rangle_\rho \approx 0.25$, $\epsilon \approx 0.01$). Then, if we obtain $\chi^2(s_0) \approx 1$, this tells us that the mean deviation of $F(M^2,s_0)$ from a constant value in the region $[M_-^2, M_+^2]$ is about $\epsilon$. To find the minimum value of $\chi^2$ and the corresponding $s_0$, we used the code Mathematica.